\documentclass[prl,showkeys,showpacs,twocolumn,aps,amsmath,amssymb,reprint,superscriptaddress]{revtex4-1}

\usepackage[dvips]{graphicx}
\usepackage{multirow}
\usepackage{array}
\usepackage{bm}
\usepackage{amsmath}
\usepackage{threeparttable}
\usepackage{epstopdf}
\usepackage{dcolumn}
\usepackage{ulem}
\usepackage[usenames,dvipsnames]{color}

\begin{document}

\title{Non-Arrhenius ionic conductivities in glasses due to a distribution of activation energies}

\author{C. Bischoff }
\author{K. Schuller}
\author{S. P. Beckman}
\author{S. W. Martin}
\email{swmartin@iastate.edu}
\affiliation{Department of Material Science and Engineering, Iowa State University, Ames, Iowa 50011, USA}%

\date{\today}

\begin{abstract}

Previously observed non-Arrhenius behavior in fast ion conducting 
glasses [\textit{Phys.\ Rev.\ Lett.}\ \textbf{76}, 70 (1996)] 
occurs at temperatures near the glass transition temperature, $T_{g}$, and is attributed to changes in the 
ion mobility  due to ion trapping mechanisms 
that diminish the conductivity
and result in a decreasing conductivity with increasing temperature.  
It is intuitive that disorder in glass 
will also result in a distribution 
of the activation energies (DAE) for ion conduction, which should increase the conductivity with 
increasing temperature, yet this has not been identified in the literature. 
In this paper, a series of high precision ionic conductivity measurements 
are reported for  
$0.5\mbox{Na}_{2}\mbox{S}+0.5[x\mbox{GeS}_{2}+\left(1-x\right)\mbox{PS}_{\frac{5}{2}}]$
glasses with compositions ranging from $0 \leq x \leq 1$.  
The impact of the cation site 
disorder on the activation energy is identified and explained using 
a DAE model.  
The absence of the non-Arrhenius behavior in other glasses is 
explained and it is predicted which glasses are expected to accentuate the 
DAE effect on the ionic conductivity.

\end{abstract}


\maketitle

It is universally accepted that the ionic conductivity of a simple solid electrolyte should obey an Arrhenius relation.  The temperature dependence of the conductivity is controlled by the activation energy, which is physically interpreted as the energy barrier that must be overcome for an ionic charge carrier to jump to an adjacent site. This simple relationship is expected in crystals where the activation energy is a singular value.  In highly disordered systems, such as ionic glasses, the ion conduction processes are expected to arise from jumps over a distribution of such energy barriers, therefore, a distribution of activation energies (DAE) is anticipated. The atomic level origin of the DAE arises from the structural disorder in the glass, at both short and long length scales, which results in wide and continuous distributions of bond distances, bond angles, and in some cases coordination numbers.
This DAE should result in the ionic conductivity having a non-Arrhenius temperature dependence, where the low energy 
barriers are crossed at low temperatures and the high energy barriers are crossed at high temperatures. 
Astonishingly, however, the vast majority of published studies of ion-conducting glasses report an Arrhenius relationship with a single activation energy. Until now it remained unknown why the chemical and structural complexity that is intrinsic to the glassy state does not result in a dramatic deviation from simple Arrhenius behavior.  In this letter we resolve this long standing question.

There are a limited number of studies observing non-Arrhenius conductivity in glasses.  One class of non-Arrhenius behavior is exemplified by the measurements of Kincs and Martin on silver halide doped sulfide glasses ~\cite{Kincs1996}. At low temperatures, the conductivity maintains a linear Arrhenius behavior, but at higher temperatures it develops a distinctly negative curvature. The models developed to explain this behavior have all focused on site-hopping diffusion and mobility ~\cite{Murugavel2005, Maass1996, Malki2006, Ngai1996, Ngai1998}. Malki \textit{et al.}\ identify a rigid-to-floppy transition, which depends on both the composition and the temperature to delineate the high-temperature, non-Arrhenius behavior from the low temperature regime ~\cite{Malki2006}. In essence, this form of non-Arrhenius behavior is caused by a temperature dependent mobility due to an ion trapping mechanism that reduces the mobility at high temperatures ~\cite{kim1996, martin2002, mei2004}.

A second class of non-Arrhenius behavior is characterized by a positive curvature of the conductivity across the entire temperature region.  Archetypal examples of this behavior are found in Namikawa's measurements of mixed-alkali glasses~\cite{Namikawa1975} and Murugavel \textit{et al.}\ for multicomponent phosphosilicate glasses~\cite{Murugavel2010}. Namikawa attributes this behavior to the existence of multiple charge carrying species and uses two straight lines to approximate the curved behavior, with each slope indicating the activation energy of one of the species.  While it is possible to attribute the positive-curvature non-Arrhenius behavior to the activation energies of two charge carriers; the same non-Arrhenius behavior is observed in glasses with single charge carriers: Murugavel and Roling for $x \mbox{Na}_{2}\mbox{O} + \left(1-x\right) \mbox{B}_{2}\mbox{O}_{3}$~\cite{Murugavel2007}, and Imre \textit{et al.}\ for $0.2 \mbox{Na}_{2}\mbox{O}  + 0.8 \mbox{B}_{2}\mbox{O}_{3}$ and $0.2 \mbox{Rb}_{2}\mbox{O} + 0.8 \mbox{B}_{2}\mbox{O}_{3}$~\cite{Imre2002}. 

In this letter we report on a series of high precision measurements on mobile Na ion containing sulfide glasses.  The data demonstrates the same positive-curvature non-Arrhenius behavior described in references~\cite{Namikawa1975, Murugavel2007, Imre2002, Murugavel2010}. Following the lead of our earlier work \cite{kim1996, martin2002, mei2004}, we develop a DAE model and demonstrate an excellent agreement between our model and the data. The results of this model allow us to explain why such non-Arrhenius behavior is not commonly seen in other glasses and how to modify a glass to enhance the influence of the DAE. Finally, DAEs have been used to describe ultrasonic attenuation (UA) measurements of glass where the UA loss peaks were observed to be wider than could be described by a single activation energy~\cite{borjesson1987,gilroy1981}.

Mixtures with compositions $0.5\mbox{Na}_{2}\mbox{S}+0.5[x\mbox{GeS}_{2}+\left(1-x\right)\mbox{PS}_{\frac{5}{2}}]$, where $x$ ranges from 0.0 to 1.0, were prepared from $0.5\mbox{Na}_{2}\mbox{S} + 0.5\mbox{GeS}_{2}$ and $0.5\mbox{Na}_{2}\mbox{S} + 0.5\mbox{PS}_{\frac{5}{2}}$. All syntheses were carried out in a high purity N$_{2}$ glove box. The mixtures were melted in covered vitreous carbon crucibles inside a tube furnace at 730 $^{\circ}$C for 3 minutes. The mixture was removed from the furnace and allowed to cool inside the crucible. The mass loss due to sublimation of reactants, typically less than one percent, was recorded to ensure compositional accuracy. Disk-shaped specimens were then prepared by remelting the glass and pouring it into disk-shaped brass molds that were held 30 $^{\circ}$C below the glass transition temperature. The glasses were transparent and had yellow to amber colors. After annealing for 30 minutes, the disks were cooled to room temperature at 1 $^{\circ}$C/min. The disks were polished and circular gold electrodes were sputtered onto both faces of the glass. A mask was used to assure that the electrodes were registered on both sides of the disk. Using a Novocontrol Technologies Concept 80 impedance spectrometer, the complex impedance spectra were measured from 0.1 Hz to 3 MHz and temperatures from -50 $^{\circ}$C to 150 $^{\circ}$C, where -50$^{\circ}$C is the lower limit of the components of the custom cell used for O$_{2}$ and H$_{2}$O sensitive materials and 150$^{\circ}$C is below the annealing temperature for all samples in this study. The temperature was held within $\pm0.5$ $^{\circ}$C of the nominal set point for three minutes prior to data collection to stabilize the temperature. The direct current (DC) conductivity was found by fitting the complex impedance arc and using the disk's thickness and electrode area.

The ionic conductivity of glass can be expressed as the Nernst-Einstein relation, a modified functional form of which is,
\begin{equation}\label{NE}
\sigma_{DC}\left(T\right)=\frac{\sigma_{0}}{T} \exp\left(\frac{-\Delta E_{a}}{RT}\right),
\end{equation}
where $R$ is the gas constant and the prefactor, $\sigma_{0}$, contains the ion charge, number density, and diffusivity. The activation energy, $\Delta E_{a}$, is an energy barrier that the charge carrying ions must overcome to conduct through the glass network. In conductivity studies, it is common practice to use the measurements to determine the values of $\sigma_{0}$  and $\Delta E_{a}$ as constant material properties. However, this assumes that both parameters are singular constant values, whereas the disordered short and intermediate range chemical structures seen by mobile ions in glasses should result in $\Delta E_{a}$ being a DAE associated with these local variations.  The simplest correction to this assumption is to replace $\Delta E_{a}$  in equation \ref{NE} with a temperature dependent expectation energy, $\langle \Delta E \rangle$, that can be found from 
\begin{equation}\label{expE}
\langle \Delta E \rangle = \int_{0}^{\infty}\Delta EP\left(\Delta E,T\right)d\Delta E,
\end{equation}
where $P\left(\Delta E,T\right)$ is the probability distribution function of the activation energies of the mobile ions at temperature T. This probability distribution introduces a temperature dependence to the expected energy and is written using the Boltzmann relation,
\begin{equation}\label{Boltz}
P\left(\Delta E,T\right)=\frac{g\left(\Delta E\right)\exp\left(\frac{-\Delta E}{RT}\right)}{\int_{0}^{\infty}g\left(\Delta E^{\prime}\right)\exp\left(\frac{-\Delta E^{\prime}}{RT}\right)d\Delta E^{\prime}}.
\end{equation}
The function $g\left(\Delta E\right)$, in equation \ref{Boltz}, is the temperature-independent probability distribution function of the activation energies, DAE, of all the ions in the glass. In this method, it is assumed that the energy landscape seen by the mobile cations is fixed upon quenching and with increasing temperature the number of these fixed energy barriers being overcome increases according to the Boltzmann distribution. For a given $g\left(\Delta E\right)$, equation \ref{Boltz} is integrated to determine the probability distribution function of the mobile ions as a function of temperature, $P\left(\Delta E,T\right)$. Subsequently, equation \ref{expE} is used to determine the temperature dependent expectation energy, $\langle \Delta E \rangle$, which then is used in place of $\Delta E_{a}$ in equation \ref{NE}.

\begin{figure}[bt]
\setlength{\unitlength}{1.0in}
\centering
\includegraphics[width=\columnwidth]{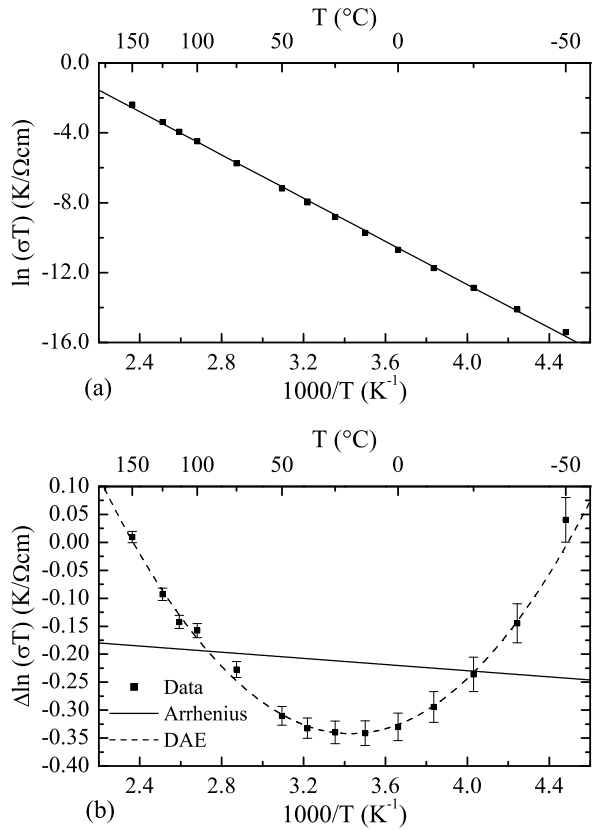}
\caption{The measured ionic conductivity for the glass composition 
$0.5\mbox{Na}_{2}\mbox{S}+0.5[0.7\mbox{GeS}_{2}+0.3\mbox{PS}_{\frac{5}{2}}]$. 
Frame (a) shows the conductivity plotted with a best-fit Arrhenius line, 
where the experimental error is smaller than the symbols.
The data in frame (b) is the difference between the conductivity 
data, from (a) 
minus a point from a line arbitrarily drawn between the highest and 
lowest data points. 
The solid line in (b) is the match between the Arrhenius fit in (a) and 
the arbitrary straight line. The dashed line is the result of the DAE model.}
\label{fig1}
\end{figure}

The measured ionic conductivity of a typical glass in this series is plotted in FIG. \ref{fig1}(a). On first inspection, the data appears to be Arrhenius; however, closer evaluation reveals a positive, upward curvature. This non-Arrhenius nature is more evident when the data are subtracted from a straight line drawn between the lowest and highest temperature data points, as shown in FIG. \ref{fig1}(b). The deviation from linearity is distinctly quadratic and this has been observed by Imre \textit{et al.}\ for $0.2 [x\mbox{Na}_{2}\mbox{O} + \left(1-x\right)\mbox{Rb}_{2}\mbox{O}] + 0.8 \mbox{B}_{2}\mbox{O}_{3}$ glasses which is plotted in FIG. 8 of reference \cite{Imre2002}. The non-Arrhenius behavior demonstrated in FIG. \ref{fig1} is systematically present for all 10 glass specimens with compositions ranging from $x=0$ to 1.

\begin{figure}[bt]
\centering
\includegraphics[width=\columnwidth]{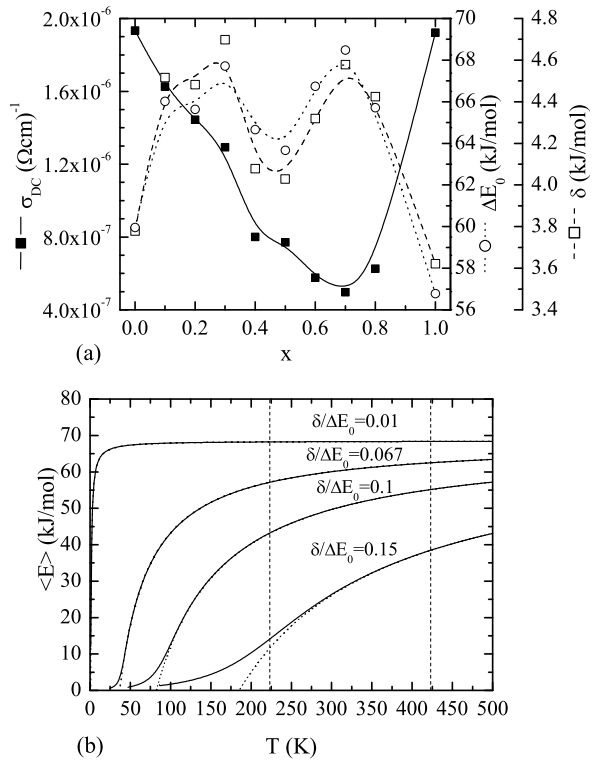}
\caption{
Frame (a) 
shows the room temperature ionic conductivity and DAE model parameters, $\Delta E_{0}$ and $\delta$, 
for glasses with composition
$0.5\mbox{Na}_{2}\mbox{S}+0.5[x\mbox{GeS}_{2}+\left(1-x\right)\mbox{PS}_{\frac{5}{2}}]$ 
as a function of composition, where the lines are guides for the eyes 
and error bars are smaller than the symbols.
Frame (b) shows the 
expectation energy as a function of 
temperature for a variety of $\delta / \Delta E_{0}$ choices. 
This family of curves asymptotes toward $\Delta E_{0}$ with increasing 
temperature. The rate of convergence depends on the 
value of $\delta / \Delta E_{0}$.  The dashed lines bracket the
 temperature range for the glass specimens studied here.
The solid lines show the exact integral, from equation \ref{expE}, 
and the dotted lines show the expectation energy from equation \ref{approx}.
The data plotted in FIG. \ref{fig1} corresponds to $x=0.7$ 
which has $\delta / \Delta E_{0} = 0.067$.}
\label{fig2}
\end{figure}

For these glasses, the DAE barriers, $g\left(\Delta E\right)$, can be 
taken to be Gaussian, with a mean value of $\Delta E_{0}$ and 
standard deviation of $\delta$.~\cite{kim1996, martin2002, mei2004}
This results in equation \ref{expE} being expressed, 
\begin{eqnarray*}
\lefteqn{\langle \Delta E\rangle = } \\
&  & \int_{0}^{\infty}E \frac{\sqrt{\frac{1}{2 \pi \delta^{2}}} \exp \left(  -\frac{\left(E - \Delta E_{0}\right)^{2}}{2 \delta^2} \right) \exp\left( -\frac{E}{RT} \right) }{\frac{1}{2}\exp\left( -\frac{\Delta E_{0}}{RT}+\frac{\delta^2}{2\left(RT\right)^{2}} \right) \mbox{erfc} \left( \frac{\delta}{\sqrt{2}RT} - \frac{\Delta E_{0}}{\sqrt{2}\delta} \right)} dE.
\end{eqnarray*}
which can be simplified to
\begin{equation}\label{approx}
\langle \Delta E\rangle = \Delta E_{0} - \frac{\delta^{2}}{RT}.
\end{equation}
In this way, the temperature-dependent $\langle \Delta E\rangle$ can be written 
in terms of the Gaussian parameters, $\Delta E_{0}$ and $\delta$.
To write the simplified expression in equation \ref{approx} 
it is required that $P\left(\Delta E,T\right)$ is zero at $T=0$.  
This approximation breaks 
down in the limit of small $T$ and large $\delta / E_{0}$. The 
glasses and temperatures used in this study, and in most others, 
are well within the operating limit of this approximation. The 
deviation between the full and simplified expressions 
is shown as dotted lines in FIG. \ref{fig2}(b). This 
clearly demonstrates that for $T\ge50$ K equation \ref{approx} 
holds for $\delta / \Delta E_{0}=0.067$.

Using this temperature dependent expectation energy in place 
of the constant activation energy in equation \ref{NE}, the 
natural log of $\sigma T$ becomes quadratic in $1/T$. For 
each composition of glass, the parameters $\Delta E_{0}$, $\delta$, 
and $\sigma_{0}$ were determined by replacing $\Delta E_{a}$ 
in equation \ref{NE} with $\langle \Delta E \rangle$ from 
equation \ref{approx} and best-fitting this expression to the 
experimental data. Results of this fitting are shown in FIG. \ref{fig2}(a). 
See Supplemental Material at [URL will be inserted by publisher] for 
a table of DAE parameter values for all of the glasses. As can be 
seen by the fitted curve in FIG. \ref{fig1}(b) the model matches the 
data extremely well.

To understand the impact of the DAE on the thermal behavior of the ionic conductivity, $\langle \Delta E \rangle$ is parametrically plotted for different $\delta / \Delta E_{0}$ ratios as a function of temperature in FIG. \ref{fig2}(b). The experimental temperature range of this study lies between the dashed lines. For this family of curves, it is observed that the temperature dependence is greater at lower temperatures and larger $\delta / \Delta E_{0}$ ratios; hence, the non-Arrhenius behavior will be more pronounced in material systems with DAEs that have a small mean activation energy, $\Delta E_{0}$, and a large width, $\delta$. Glass compositions with large $\Delta E_{0}$ values such as the most commonly studied silicate, borate, and phosphate oxide glasses are not expected to exhibit the non-Arrhenius behavior shown in FIG. \ref{fig1}, even if the $\delta$ is equivalent to the values found in analogous non-oxide glass systems, such as the glasses in this study.  The non-Arrhenius behavior will be enhanced in highly conductive materials with low $\Delta E_{0}$ values. For example, our model predicts that glasses that have Li as the charge carrying species will demonstrate a stronger DAE non-Arrhenius behavior than the other alkali ions because the small size of the Li ions results in a lower mean activation energy, $\Delta E_{0}$. An example of this can be found in FIG. 4 of reference \cite{Ohtomo2005} where the ionic conductivity of 
$36\mbox{Li}_{2}\mbox{S} \cdot 18\mbox{P} \cdot 46\mbox{S}$
is shown to have a highly non-Arrhenius conductivity below the T$_{g}$.

At lower temperatures, there is a stronger temperature dependence of the expectation energy because only the low energy tail of $g\left(\Delta E\right)$ contributes to $P\left(\Delta E,T\right)$ via equation \ref{Boltz}. As the temperature increases, the expectation energy asymptotes to $\Delta E_{0}$ because more of the energies in the $g\left(\Delta E\right)$ distribution begin to participate; at high temperatures the expectation energy becomes constant and the measured conductivities will appear Arrhenius. This implies that low-temperature experiments will accentuate the observation of the DAE. The exact experimental range needed is dependent on the material.

The non-Arrhenius behavior also may be enhanced by the processing methods used to prepare and the chemistries of the glass, both of which can be used to increase $\delta$. Rapid quenching of a glass melt will freeze in structures with a higher fictive temperature and, thus, a higher configurational entropy. RF-sputtered glassy thin films may also lead to a wider distribution of atomic level disorder. Additional disorder, and possibly coordinative defects, can be introduced by mechanical means such as high energy ion implantation or planetary milling. Glasses with greater chemical complexity will also lead to increased $\delta$. A distribution of anionic sites contributes to the DAE, and quench rates are known to strongly influence speciation in glasses \cite{wu2010}. Further, the use of multiple charge carriers will increase the width of the DAE, and the quadratic deviation from Arrhenius behavior observed in reference \cite{Imre2002} indicates that the existence of multiple charge carriers is not a simple matter of having two singular valued activation energies as Namikawa postulates \cite{Namikawa1975}. If only two activation energies exist, then the deviation from linearity should have a simple ``V'' shape instead of being quadratic.

Glasses with more than one glass former, such as the mixed glass former glasses in this study, should also have a wider DAE. FIG. \ref{fig2}(a) shows the compositional dependence of the ionic conductivity, $\Delta E_{0}$, and $\delta$ in the $0.5\mbox{Na}_{2}\mbox{S}+0.5[x\mbox{GeS}_{2}+\left(1-x\right)\mbox{PS}_{\frac{5}{2}}]$ glasses. As is expected, the composition  with the largest $\Delta E_{0}$, $x = 0.7$, has the lowest conductivity. The structural origins of the compositional dependence of $\Delta E_{0}$ have so far not been fully identified, but the compositional dependence of $\delta$ can be linked to the distribution of local structures. See Supplemental Material at [URL will be inserted by publisher] for the relative abundance of short-range structures in the $0.5\mbox{Na}_{2}\mbox{S}+0.5[x\mbox{GeS}_{2}+\left(1-x\right)\mbox{PS}_{\frac{5}{2}}]$ glasses. Structures are labeled Q$^{\mbox{z}}$, where Q denotes the glass-forming cation (P or Ge) and z denotes the number of bridging sulfurs associated with the structural unit. When $x = 0$, the dominating structural unit is the P$^{1}$ group, nominally Na$_{2}$PS$_{\frac{7}{2}}$. Addition of Ge$^{2}$ groups, nominally Na$_{2}$GeS$_{3}$, leads to formation of P$^{0}$ and Ge$^{3}$ groups arising from a disproportionation reaction
\begin{equation}
P^{1} + Ge^{2} \rightarrow P^{0} + Ge^{3}.
\end{equation}
The consequence of this proposed reaction is that Na ions experience a wider distribution of chemical environments. 
Qualitatively, ternary compositions appear to have a wider distribution of structures
where no Q$^{\mbox{z}}$ group is dominant. For example, the $x = 0.3$ composition is comprised of 20.6\% P$^{1}$, 16.4\% P$^{1P}$, 32.1\% P$^{0}$, 29.6\% Ge$^{3}$, and has the largest $\delta$ value of all of the glasses.

In this letter, we have examined the impact that the DAE has on the ionic conductivities in glasses by performing a series of high precision measurements on a Na containing mixed-glass former sulfide glasses and constructing a model that shows how the DAE will affect the Nernst-Einstein relation. Using this model, we have addressed the question of why the positive-curvature non-Arrhenius conductivity associated with the DAE has only been identified very infrequently. It is significant that of the many thousands of ionic glass systems studied only a small number have reported this behavior. The most studied glasses are the more poorly conducting, yet most easily prepared, oxide glasses whose large $\Delta E_{0}$ and small $\delta$ values produce essentially an Arrhenius conductivity. It is only in materials such as the more highly conducting non-oxide glasses, especially the complex mixed-glass former glasses, that the $\Delta E_{0}$ is sufficiently low and the $\delta$ is large enough that the non-Arrhenius conductivity is apparent. This study is the first to understand and report the underlying cause of this apparent unknown behavior.

The authors gratefully acknowledge funding from the National Science Foundation. SWM and CB are funded through grant DMR-0710564 and SPB is funded through DMR-1105641.


%

\end{document}